# Calorimetric evidence for the existence of an intermediate phase between the ferroelectric nematic phase and the nematic phase in the liquid crystal RM734.

J. Thoen[1], G. Cordoyiannis[2], E. Korblova[3], D. M. Walba[3], N. A. Clark[4], W. Jiang[5], G. H. Mehl[5], and C. Glorieux[1]

[1]*KU Leuven, Department of Physics and Astronomy, Laboratory for Soft Matter and Biophysics, Celestijnenlaan 200D, 3001 Leuven, Belgium*
[2]*Condensed Matter Physics Department, Jožef Stefan Institute, 1000 Ljubljana, Slovenia*
[3]*Department of Chemistry and Soft Materials Research Center, University of Colorado, Boulder, CO, USA*
[4]*Department of Physics and Soft Materials Research Center, University of Colorado, Boulder, CO, USA*
[5]*Department of Chemistry, University of Hull, Hull HU6 7RX, United Kingdom*

**Abstract**

The idea that rod-like molecules possessing an electric dipole moment could exhibit a ferroelectric nematic phase was suggested more than a century ago. However, only recently such a phase has been reported for two quite different liquid crystals: RM734 (4-[(4-nitrophenoxy)carbonyl)]phenyl 2,4-dimethoxybenzoate) and DIO (2.3′,4′,5′-tetrafluoro[1,1′-biphenyl]-4-yl 2.6-difluoro-4-(5-propyl-1,3-dioxan-2-yl) benzoate). For RM734 a direct ferroelectric nematic ($N_F$) to classical nematic $N$ transition was reported, whereas for DIO an intermediate phase $N_x$ was discovered between the $N_F$ and the $N$ phases. Here we present high-resolution calorimetric evidence that an intermediate $N_x$ phase also exists in RM734 along a narrow temperature range between the $N_F$ and the $N$ phases.

## I INTRODUCTION

Liquid crystals (LCs) exhibit a large number of phases with different degrees of orientational and partial positional order between the liquid and the crystalline solid state [1]. For rod-like, thermotropic liquid crystals, the least ordered phase is the nematic phase, in which the molecules exhibit long-range orientational order but no long-range positional order. However, several modifications of nematic order have been observed. Chiral LC molecules form a twisted nematic (so-called cholesteric) phase and, for higher chirality, they may also exhibit blue phases forming lattices of defect lines. Transitions between uniaxial and biaxial nematic phases [2] as well as between two uniaxial nematic phases with short-range smectic order have been observed [3]. Also, nematic-nematic phase transitions were discovered for nonchiral dimer molecules [4,5].

More than a century ago Debye [6] and Born [7] argued that the presence of molecular dipoles could result in long-range ferroelectric nematic type of order. Such a phase has never been observed until recently. In two independent reports, published in 2017, Mandle *et al.* [8] and Nishikawa *et al.* [9] reported the discovery of a new type of nematic phase for two different molecules. These two compounds are referred to as RM734 (4-[(4-nitrophenoxy)carbonyl)]phenyl 2,4-dimethoxybenzoate) [8] and DIO (2.3′,4′,5′-tetrafluoro[1,1′-biphenyl]-4-yl 2.6-difluoro-4-(5-propyl-1,3-dioxan-2-yl) benzoate) [9]. The strongly polar compound RM734 was reported to exhibit two nematic phases that are separated by a first-order transition [10]. Chen *et al.* [11] presented evidence for ferroelectricity for the low-temperature nematic phase ($N_F$) of RM734, and normal nematic uniaxiality for the high-temperature one. DIO exhibits in addition to the low temperature $N_F$ phase an intermediate phasesituated between the $N_F$ and the classical nematic phase $N$ [9]. The true nature and the name for the intermediate phase is still a matter of extensive debate. This phase was initially called $M_2$ [9], and subsequently $N_x$ [12], $SmZ_A$ [13], and $N_S$ [14], considering either a smectic type of order with the molecular axis perpendicular to the layer normal or a splay nematic phase, respectively. Because calorimetry cannot render information on the structure of phases, and in view of the above variations in the nomenclature of the intermediate phase, and being consistent with our previous paper [15] we henceforth use $N_x$ to indicate the phase between the $N$ and the $N_F$ phases.

In high-resolution adiabatic scanning calorimetry (ASC) measurements (carried out in 2020) of the $N$ - $N_F$ transition of RM734, the order of the transition as well as the pretransitional fluctuation-induced critical behavior was investigated in detail [16]. The $N$ - $N_F$ transition was found to be very weakly first order with a latent heat $L = 0.115 \pm 0.005$ J/g. Power law analysis of the specific heat capacity in both phases resulted in an effective critical exponent $\alpha = 0.50 \pm 0.05$. This value and the small latent heat suggest the $N$ - $N_F$ the phase transition in RM734 to be close to a tricritical point. In a paper submitted in 2021 and published in 2022, Chen *et al.* [17] studied binary mixtures of DIO and RM734 by means of optical microscopy, differential scanning calorimetry, polarization measurements and SAXS experiments and showed complete miscibility of the isotropic phase, and the $N$ and $N_F$ phases. The $N_x$ phase present in DIO persisted only in mixture with concentrations above 40% of DIO, indicating a possible triple point around that concentration. In the second half of 2021 and the first half of 2022 some of us carried out detailed high-resolution adiabatic scanning calorimetry measurements on DIO and of binary mixtures with RM734 for concentrations above 10 % weight fraction of DIO. Contrary to the results reported in Ref. [17] we found that the intermediate $N_x$ phase was present for all studied mixtures, albeit with a strongly decreasing temperature range with decreasing concentration of DIO. Both the $N$ - $N_x$ and the $N_x$ - $N_F$ transitions were found to be first order ones, with very small latent heats [15]. Since high-resolution data existed [16] for RM734, we did not remeasure it at that time. However, in a Physical Review E Perspective paper, Sebastián *et al.* [14] conjectured the possibility of the existence of a very narrow $N_x$ phase also in RM734. Their conjecture was based on the observation of freezing of the nematic fluctuations in a very small temperature range in the $N$ phase close to $N_S$, similar to what was observed for $N_x$ in DIO. Although, in our previous work on RM734 we did not observe any thermal signature of a possible $N$ - $N_x$ transition [16], we have made further efforts to improve the resolution of the measurements and revisit the RM734 problem. In this paper, we present new ASC results obtained for 20 scanning (heating and cooling) runs on 5 RM734 samples from 4 different batches.

## II EXPERIMENT

### A. Adiabatic scanning calorimetry (ASC)

Adiabatic scanning calorimetry (ASC) yields simultaneously and continuously the temperature evolution of the heat capacity $C_p$ and the enthalpy $H$ of a sample [18-21]. Contrary to all other calorimetric methods operating with a constant scanning rate, in ASC a constant heating or cooling power is applied to a sample holder containing the sample. This holder with sample is placed inside a surrounding adiabatic shield. The adiabatic conditions are maintained by means of a number of shields surrounding the sample holder and sample, as well as by pumping the air from the calorimeter.

Detailed descriptions of the 'classical' ASC implementations and results can be found in [18-21] and references therein. A problem in the older types of ASCs lies in maintaining adiabatic conditions over long time and wide temperature ranges. In the past decade the development of the novel Peltier-element-based adiabatic scanning calorimeter (pASC) has largely eliminated these issues, with greater user-friendliness and much wider temperature ranges as well as much smaller samples (typically a few tens of milligram) [22-24].

In a pASC heating run the heat exchange between the inner shield and the sample holder is cancelled by keeping the temperature difference, measured by a Peltier element, zero at all times, using a proportional-integrating (PI) control loop. In a cooling run, the PI-control keeps a non-zero voltage offset, allowing to control and monitor the heat exchange. During any run, the sample temperature $T(t)$ is recorded as a function of time $t$. Together with the applied power $P$ this directly results in the enthalpy curve

$$H(T) - H(T_0) = \int_{t_0}^{t(T)} P dt = P[t(T) - t_0(T_0)], \quad (1)$$

where $H(T_0)$ is the enthalpy of the system at temperature $T_0$ at the starting time $t_0$ of the run. The heat capacity $C_p(T)$ is then calculated from the ratio of the known constant power $P$ and the changing temperature rate $\dot{T} = dT/dt$,

$$C_p = \frac{P}{\dot{T}}. \quad (2)$$

The specific heat capacity $c_p(T)$ and the specific enthalpy $h(T)$ are obtained by using the sample mass and the calibrated background values of the empty calorimeter and of the used sample cells. Keeping $P$ constant in Eq. (2) is completely opposite to the operation of a differential scanning calorimeter, where a constant rate $\dot{T}$ is imposed on a sample and on a reference and where the difference in heat flux, $\Delta P(t)$, between the sample and the reference, is measured.

### B. Materials and samples

The samples 1-UCO and 2-UCO named in Table 1 were both from the same batch of RM734 synthesized at the Soft Materials Research Center of the University of Colorado, USA (UCO). The samples 1-HULL-2, 1-HULL-3, 2-HULL-3, and 3-HULL-3 were from two different batches of RM734 (HULL-2 and HULL-3), synthesized at the Organic and Materials

Chemistry Group of the Chemistry Department of the University of Hull, UK (HULL). RM734 material from batch HULL-1 has been used in preparing mixtures with DIO [14]. The samples 1-INSTEC and 2-INSTEC were from RM734 material purchased from INSTEC Inc, Boulder, CO, USA (INSTEC). All compounds were used as received at the Laboratory for Soft Matter and Biophysics, Department of Physics and Astronomy, KU Leuven, Belgium. The samples (typically from 30 to 60 mg) were transferred into stainless steel sample holders (Mettler Toledo 120 µl medium pressure crucible) and hermetically sealed. These crucibles are vacuum-tight-closed with a miniature elastic O-ring between the cell body and the lid.

**III RESULTS AND DISCUSSION**

   A. **Specific heat capacity**

Similar to classical ASC, pASC measurements also yield simultaneously the temperature dependence of the specific enthalpy $h(T)$ and of the specific heat capacity $c_p(T)$. However, in looking at small features for a possible $N$ - $N_x$ transition temperature, it turned out that possible small variations in $h(T)$ near the expected transition were largely obscured by the large non-linear temperature-dependent pretransitional background related to the transition to the $N_F$ phase. Therefore, in the present study, we have focused on $c_p(T)$ which is the temperature derivative of $h(T)$, and more sensitive to effects that occur over a narrow temperature range.

In Table 1, an overview of the different (heating and cooling) runs that are presented further on is given. Apart from run 1 (measured in August 2020), published in Ref. [16], all other runs in the table correspond to more recent efforts to look for convincing evidence of an $N$ - $N_x$ transition with settings of the calorimeter that were aimed at improved resolution and stability. In Table 1, the first column is the run number and the second column is the date of the measurement. The third column gives the identifying name of the sample measured in a given run. In the name of the sample, the digit in front gives the sample number of a given batch. The acronyms stand for the origin of the material (see section II.B). The digit after the acronym gives the number of the batch where appropriate. Further columns in the table give additional information regarding the sample mass, the type of the run (heating or cooling), an average value for the scanning rate, the observed values for the $N_F$ - $N_x$ and $N_x$ - $N$ transition temperatures, and the temperature width of the $N_x$ phase. Regarding the rates, it should be noted that these are only average values, because in ASC runs the applied (heating or cooling) power is kept constant and the temperature dependent rate (slowing down at phase transitions) is measured (see Eq. 2).

In Fig. 1, the original data from Ref. [16] are compared with the results of a cooling run of a second sample of the same batch of RM734. As indicated by the arrow, there is a small peak visible about 1 °C above the $N_F$ - $N_x$ transition temperature. The existence of this feature is confirmed in Fig. 2 for a cooling run (run 10) as well as for a heating run (run 12) for the same sample in the same sample cell as measured in Ref. [16]. Thus, the indication for the existence of an intermediate $N_x$ phase between the $N_F$ and the $N$ phases is present in two different samples of the same (UCO) RM734 batch. Figures with data from different runs on the same samples can be found in the Supplementary Material. Fig. 3 depicts the results of a cooling and a heating run for two different samples from batch HULL-3. In both cases a very small peak points at the presence of a $N_x$ - $N$ transition again about 1 °C above the $N_F$ - $N_x$ transition. This is confirmed in Fig. 4 by a cooling run from another sample from the same batch and a cooling run for a sample from the HULL-2 batch. The run for the latter sample was special in the sense that it was not obtained upon cooling with the cooling power kept constant, but in free cooling with the sample initially at 145 °C and the first (adiabatic) shield kept at a constant temperature of 100 °C. More recently we purchased RM734 from the commercial supplier INSTEC Inc (USA) and performed additional pASC measurements. The $c_p(T)$ results for both runs are given in Fig. 5. Here evidence for a weak $N_x$ - $N$ transition is also present. In the Supplementary Material $c_p(T)$ figures for the other runs not discussed here can be found.

### B. Transition temperatures and width of the $N_x$ phase

In columns 7 and 8 of Table 1 the transition temperatures (maxima of the $c_p(T)$ peaks) of the $N_F$ - $N_x$ and of the $N_x$ - $N$ transitions are given. As can be seen in Fig. 6, quite substantial differences are observed between the different batches of the RM734 material. The transition temperatures fall in the temperature range between 126 °C and 132 °C. In the 9$^{th}$ column of Table 1 also the widths, $\Delta T$, of the $N_x$ phase observed for the different runs are listed. Here, and in Fig. 6, no clear batch dependence can be observed. The $\Delta T$ values vary between 1.02 K and 1.24 K. This results in an average value of $\Delta T = 1.15 \pm 0.10$ K for the width of the $N_x$ phase of RM734. In the phase diagram shown in Fig. 8 of Ref. [15] for mixtures of DIO and RM734, in addition to the reported data points for the transition temperatures, also quadratic trendlines through the $N_F$ - $N_x$ data points and the $N_F$ - $N_x$ ones, are displayed. Extrapolation of these two trend lines to zero DIO concentration results in a width for $N_x$ in RM734 of $\Delta T = 0.89$ K, which is close to the above average value.

### C. Transition heat of the $N_x$-$N$ transition

In the figures discussed so far, and for the temperature ranges where the two transitions $N_F$ - $N_x$ and $N_x$ - $N$ are present, the $c_p(T)$ peak for $N_x$ - $N$ is extremely small in comparison with the one for $N_F$ - $N_x$. The visibility of the $N_x$ - $N$ peak can be substantially improved by reducing the $c_p(T)$ and $T$ range around the transition. This is done in the top part of Fig. 7 for the heating run 12, where the peak is clearly visible on top of the substantial pretransitional background increase (with decreasing temperature) associated with the $N_F$ - $N_x$ transition. From this figure it is also possible to derive the transition heat $\Delta h$ of the $N_x$ - $N$ transition by integrating the part of the experimental curve above the background given by the solid curve. This curve is a fit to the $c_p(T)$ data between 131.5 and 132.5 °C, excluding the range corresponding with the $N_x$ - $N$ transition. The corresponding integration result is displayed in the lower part of Fig. 7. From careful inspection of this curve, it can be concluded that the specific enthalpy change $\Delta h$ needed to convert from the $N_x$ to the $N$ phase is $\Delta h = 0.0045 \pm 0.0005$ J/g. Similar analyses for other runs gave similar values for $\Delta h$. It should be noted that $\Delta h$ includes the true latent hear $L$ and the pretransition enthalpy increase $\delta h$. Based on comparison with DIO the value of $\Delta h$ is most likely equal to the latent heat. This value is extremely small in comparison with $L = 0.115 \pm 0.005$ J/g for the $N_F$ - $N_x$ transition and 2 J/g typical for a normal isotropic-nematic transition. In Fig.10 of Ref. [15] for mixtures of DIO and RM734, in addition to the reported data points for the latent heats, also quadratic trendlines through the $N_F$ - $N_x$ data points and the $N_x$ - $N$ ones, are displayed. Extrapolation of the trend line of $N_x$ - $N$ to zero DIO concentration results in a value of $L$ = 0.0042 J/g, which is quite close the value L = 0.0045 ± 0.0005 J/g derived from Fig. 7.

## IV CONCLUSIONS

Although in our previous work on RM734 we did not observe in our data a thermal signature for a possible $N$ - $N_x$ transition, we decided to make further efforts to improve the resolution and accuracy of the measurements and revisited the RM734 problem by focusing on the $c_p(T)$ profiles of various slow scans on samples from different batches. In this paper we presented new pASC results obtained for 20 scanning (heating and cooling) runs on five RM734 samples from four different batches as identified in Table 1. The results of 10 of these runs are displayed and discussed in Figs. 1 – 5. The results for the remaining runs are given in the associated Supplementary Material. From the analysis of all these results it is concluded that also RM734 exhibits an intermediate $N_x$ phase between the ferroelectric nematic $N_F$ phase and the normal nematic phase $N$. For the temperature width of the $N_x$ phase, a value of $\Delta T$ = 1.15 ± 0.10 K is obtained. For the $N_x$ - $N$ transition, a heat of transition $\Delta h$ = 0.0045 ± 0.0005 J/g was found.

## ACKNOWLEDGMENTS

G.C. acknowledges the financial support of Project P1-0125 of the Slovene Research Agency.

**Supplementary Material**

……………..

Calorimetric evidence for the existence of an intermediate phase between the ferroelectric nematic phase and the nematic phase in the liquid crystal RM734

_________________________________________________________________________________

Calorimetric evidence for the existence of an intermediate phase between the ferroelectric nematic phase and the nematic phase in the liquid crystal RM734

___

**Table 1.** Identification and results for all the runs carried out in search of the existence of a $N_x$ - $N$ transition in RM734. From left to right, the number and date of run, the origin and the mass of sample, the type of run (heating or cooling), the average scanning rate, the transition temperatures $T_{N_F-N_x}$ and $T_{N_F-N_x}$, as well as the range $\Delta T$ of the $N_x$ phase are presented.

| run # | date | sample | mass (mg) | run type | average rate (Kmin$^{-1}$) | $T_{N_F-N_x}$ (°C) | $T_{N_x-N}$ (°C) | $\Delta T$ (K) |
|---|---|---|---|---|---|---|---|---|
| 1 | 2020.08.02 | 1-UCO | 60.1 | heating | 0.0301 | 131.65 | | |
| 2 | 2022.11.04 | 2-UCO | 33.4 | cooling | -0.0375 | 131.09 | 132.33 | 1.24 |
| 3 | 2022.11.03 | 2-UCO | 33.4 | cooling | -0.0259 | 131.03 | 132.20 | 1.17 |
| 4 | 2022.10.31 | 2-UCO | 33.4 | cooling | -0.0209 | 130.94 | 132.08 | 1.14 |
| 5 | 2022.11.01 | 2-UCO | 33.4 | heating | 0.0095 | 130.93 | 132.06 | 1.19 |
| 6 | 2022.11.03 | 2-UCO | 33.4 | heating | 0.0192 | 131.05 | | |
| 7 | 2022.10.29 | 2-UCO | 33.4 | heating | 0.0385 | 131.14 | 132.32 | 1.18 |
| 8 | 2022.10.31 | 2-UCO | 33.4 | cooling | -0.0161 | 131.06 | 132.29 | 1.23 |
| 9 | 2022.11.07 | 1-UCO | 60.6 | cooling | -0.0187 | 130.88 | 132.06 | 1.18 |
| 10 | 2022.11.10 | 1-UCO | 60.6 | cooling | -0.0235 | 131.13 | 132.33 | 1.2 |
| 11 | 2022.11.08 | 1-UCO | 60.6 | heating | 0.0083 | 131.10 | 132.31 | 1.21 |
| 12 | 2023.03.20 | 1-UCO | 60.6 | heating | 0.0345 | 130.89 | 132.11 | 1.22 |
| 13 | 2023.03.21 | 1-UCO | 60.6 | cooling | -0.0656 | 130.86 | 132.08 | 1.22 |
| 14 | 2022.12.25 | 1-HULL-2 | 29.3 | free cooling | -0.0909 | 127.37 | 128.39 | 1.02 |
| 15 | 2022.03.28 | 1-HULL-3 | 28.9 | cooling | -0.0707 | 128.34 | 129.38 | 1.04 |
| 16 | 2023.03.28 | 2-HULL-3 | 28.9 | heating | 0.0399 | 128.43 | 129.51 | 1.08 |
| 17 | 2023.03.29 | 3-HULL-3 | 28.9 | heating | 0.0199 | 128.34 | 129.37 | 1.03 |
| 18 | 2023.11.28 | 1-INSTEC | 27.8 | cooling | -0.0287 | 126.79 | 127.91 | 1.12 |
| 19 | 2023.12.22 | 2-INSTEC | 65.9 | cooling | -0.0238 | 126.26 | 127.34 | 1.08 |
| 20 | 2023.12.26 | 2-INSTEC | 65.9 | cooling | -0.0246 | 126.34 | 127.53 | 1.19 |

Calorimetric evidence for the existence of an intermediate phase between the ferroelectric nematic phase and the nematic phase in the liquid crystal RM734
___________________________________________________________________________

**Figure captions**

FIG. 1. (Color online). Temperature dependence of the specific heat capacity $c_p(T)$ of the ferroelectric nematic liquid crystal compound RM734. (a) Data from the heating run 1 for sample 1-UCO from Table 1 and Ref. [16]; (b) Data from cooling run 2 for sample 2-UCO from Table 1. The arrow indicates the location of the $N_x$-$N$ transition.

FIG. 2. (Color online). Temperature dependence of the specific heat capacity $c_p(T)$ of the ferroelectric nematic liquid crystal compound RM734. (a) Data from the cooling run 10 for sample 1-UCO from Table 1; (b) Data from heating run 12 for sample 1-UCO from Table 1. The arrow indicates the location of the $N_x$-$N$ transition.

FIG.3. (Color online). Temperature dependence of the specific heat capacity $c_p(T)$ of the ferroelectric nematic liquid crystal compound RM734. (a) Data from the cooling run 15 for sample 1-HULL-3 from Table 1; (b) Data from heating run 16 for sample 2-HULL-3 from Table 1. The arrow indicates the location of the $N_x$-$N$ transition.

FIG. 4. (Color online). Temperature dependence of the specific heat capacity $c_p(T)$ of the ferroelectric nematic liquid crystal compound RM734. (a) Data from the heating run 17 for sample 3-HULL-3 from Table 1; (b) Data from free cooling run 14 for sample 1-HULL-2 from Table 1. The arrow indicates the location of the $N_x$-$N$ transition.

FIG. 5. (Color online). Temperature dependence of the specific heat capacity $c_p(T)$ of the ferroelectric nematic liquid crystal compound RM734. (a) Data from the cooling run 18 for sample 1-INSTEC from Table 1; (b) Data from cooling run 20 for sample 2-INSTEC from Table 1. The arrow indicates the location of the $N_x$-$N$ transition.

FIG. 6. (Color online). Transition temperatures of the $N_F$-$N_x$ (open circles) and $N_x$-$N$ (closed circles) transitions of RM734 as obtained from the pASC measurements of the different runs of Table 1.

FIG. 7. (Color online). (a) Temperature dependence of the specific heat capacity $c_p(T)$ of the ferroelectric nematic liquid crystal compound RM734 in a narrow temperature range around the $N_x$-$N$ phase transition temperature. (b) Enthalpy increase, associated with transition, derived from the integration of the difference between the experimental $c_p(T)$ data and the solid background curve in (a).

Calorimetric evidence for the existence of an intermediate phase between the ferroelectric nematic phase and the nematic phase in the liquid crystal RM734

___

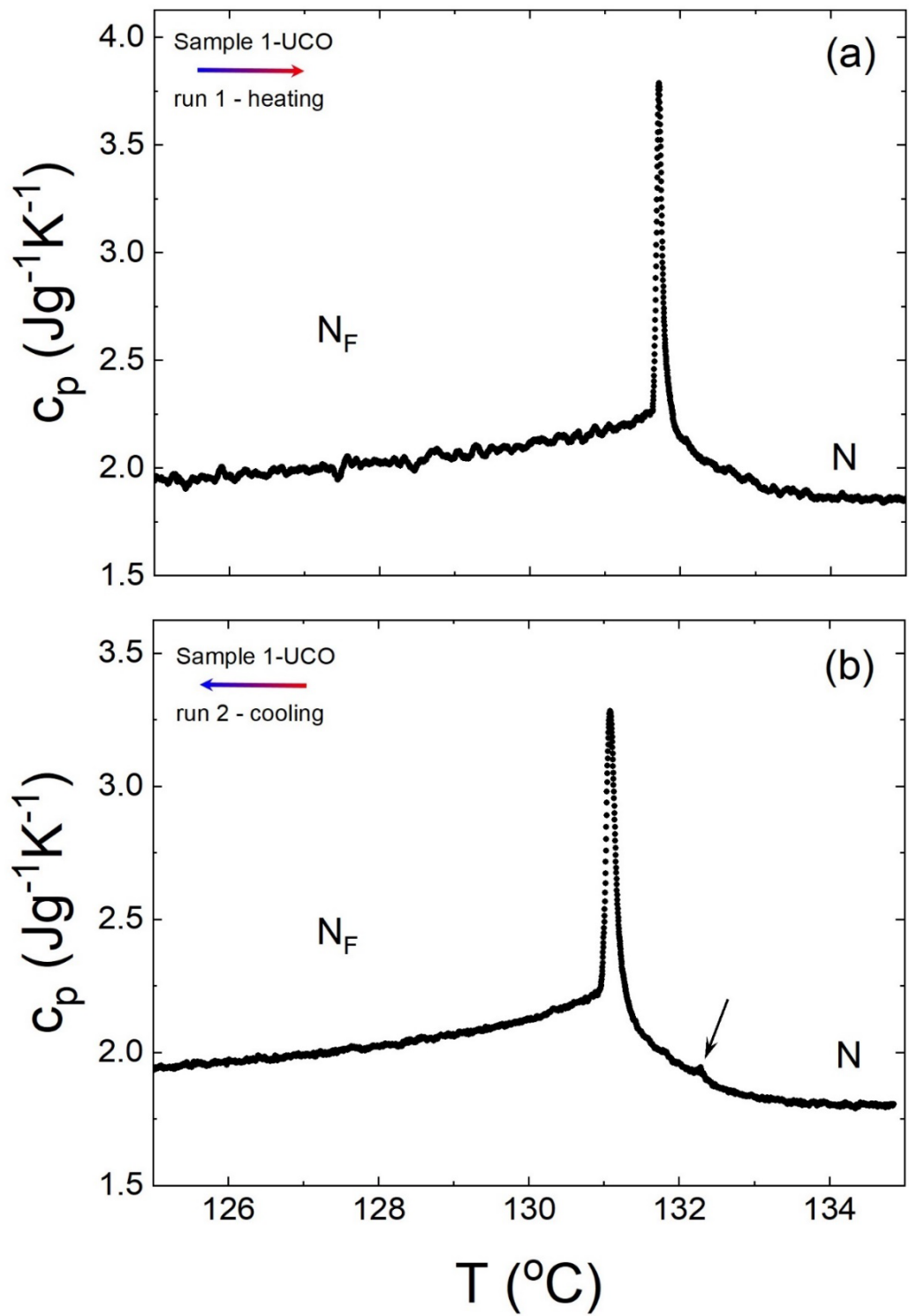

FIG. 1

Calorimetric evidence for the existence of an intermediate phase between the ferroelectric nematic phase and the nematic phase in the liquid crystal RM734

___

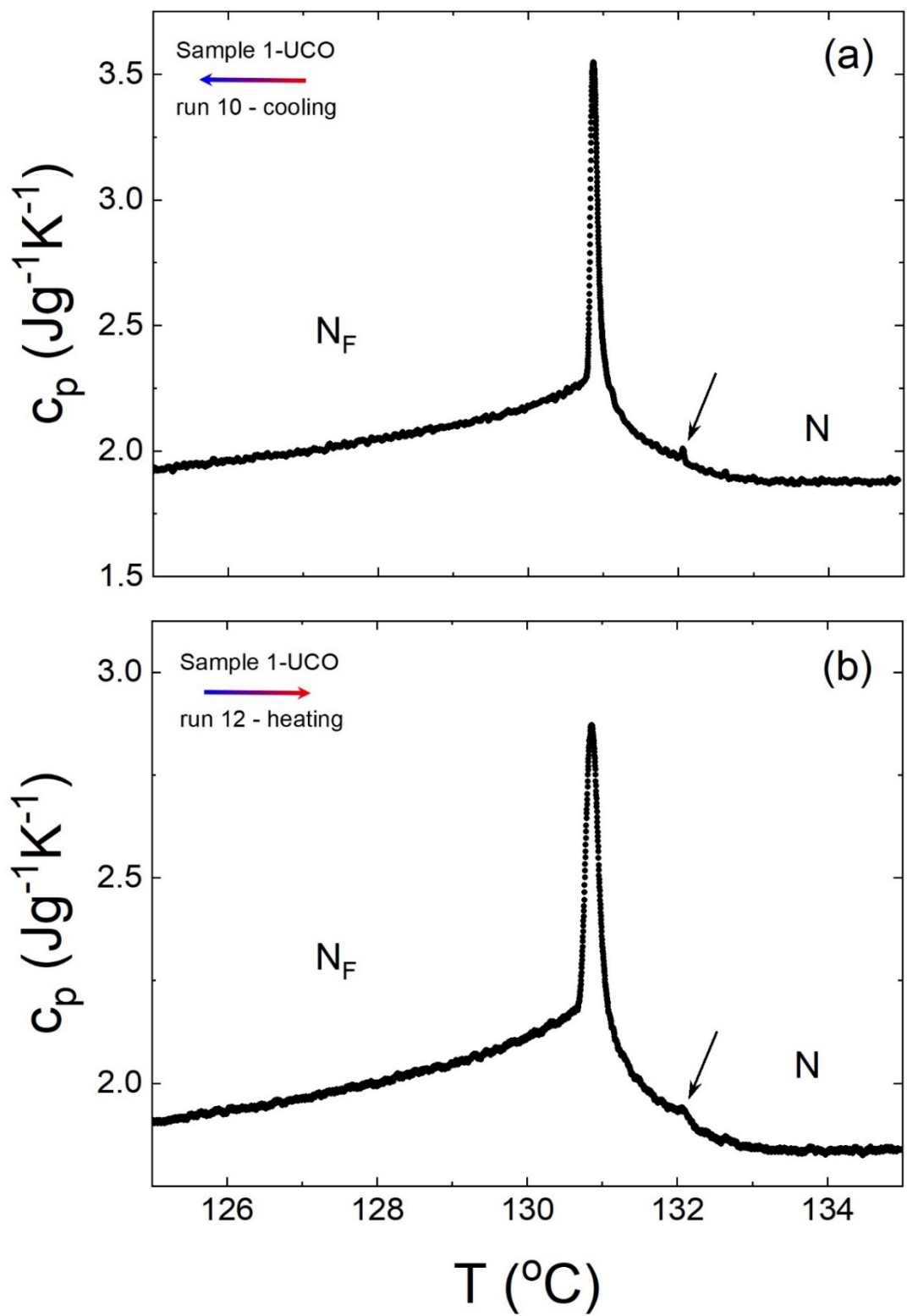

FIG. 2

Calorimetric evidence for the existence of an intermediate phase between the ferroelectric nematic phase and the nematic phase in the liquid crystal RM734

___

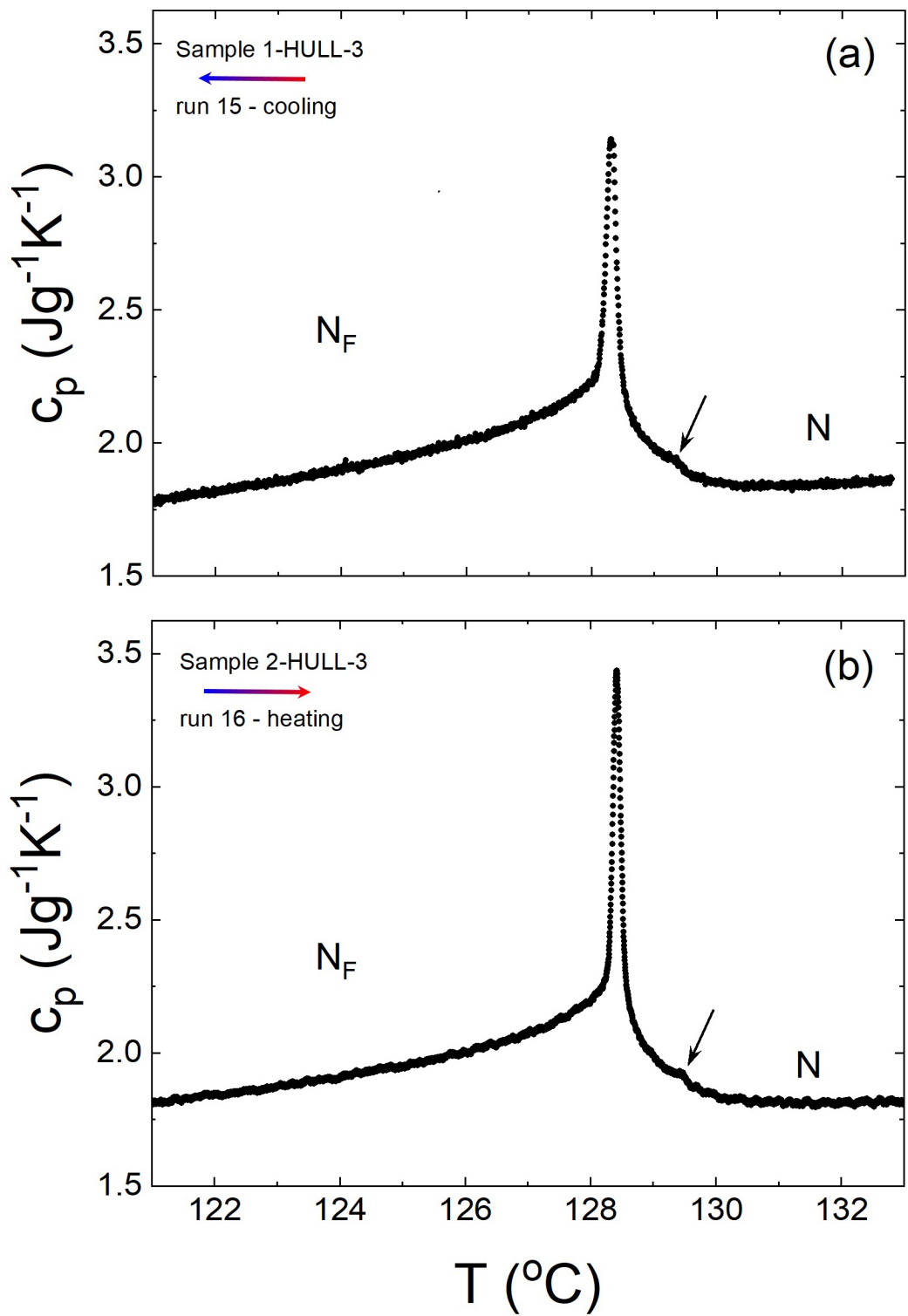

FIG. 3

Calorimetric evidence for the existence of an intermediate phase between the ferroelectric nematic phase and the nematic phase in the liquid crystal RM734

___

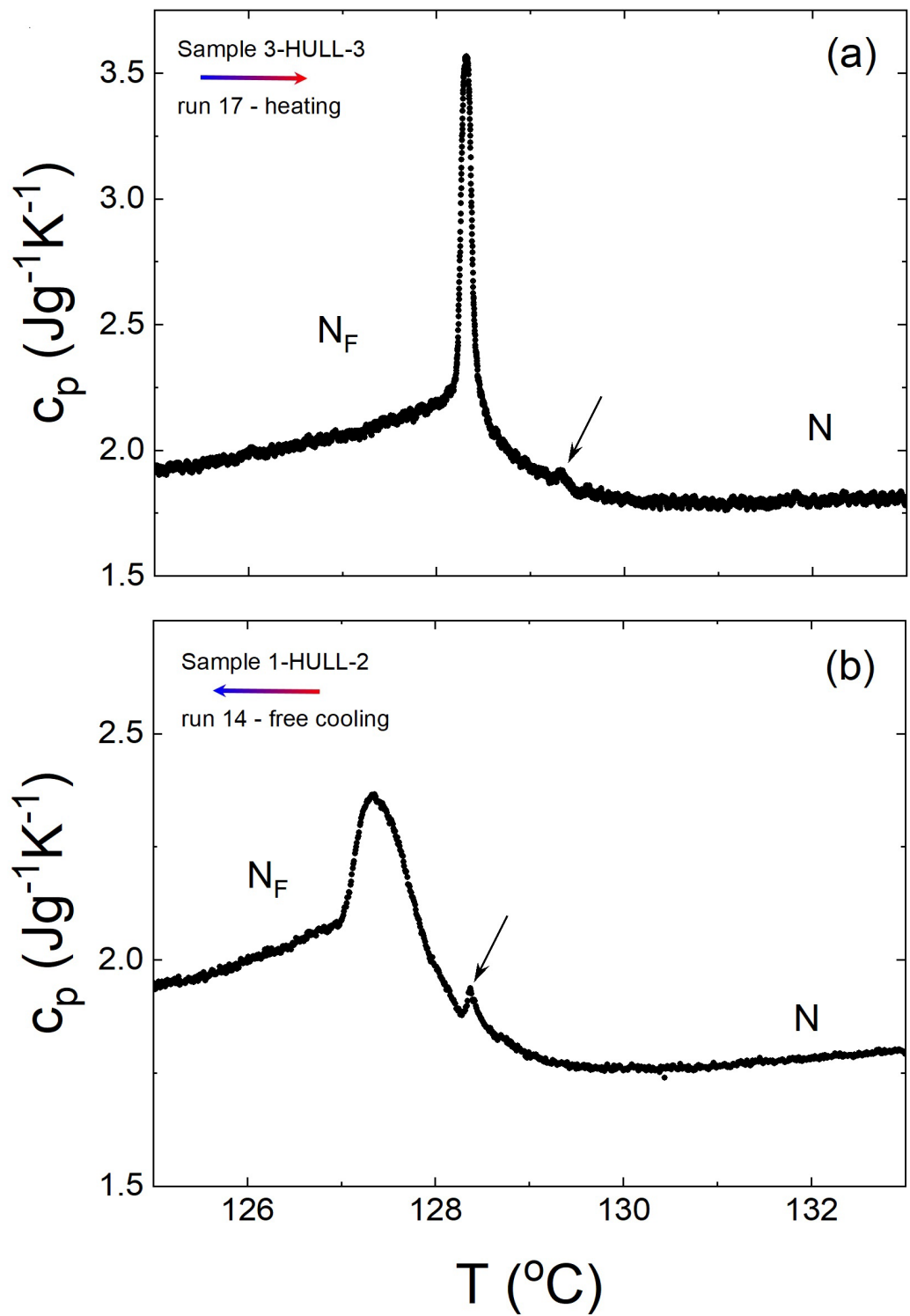

FIG. 4

Calorimetric evidence for the existence of an intermediate phase between the ferroelectric nematic phase and the nematic phase in the liquid crystal RM734

___

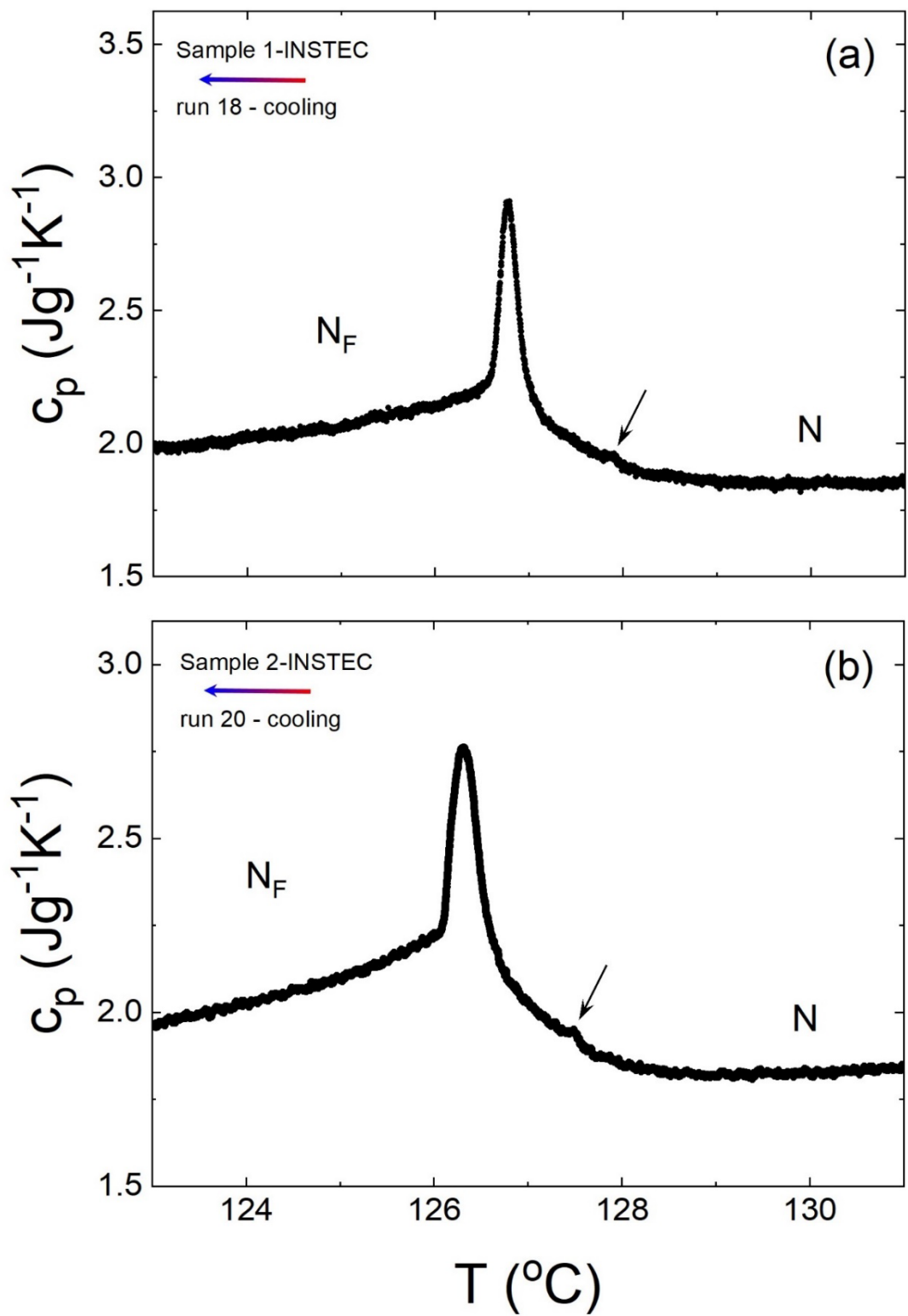

FIG. 5

Calorimetric evidence for the existence of an intermediate phase between the ferroelectric nematic phase and the nematic phase in the liquid crystal RM734

---

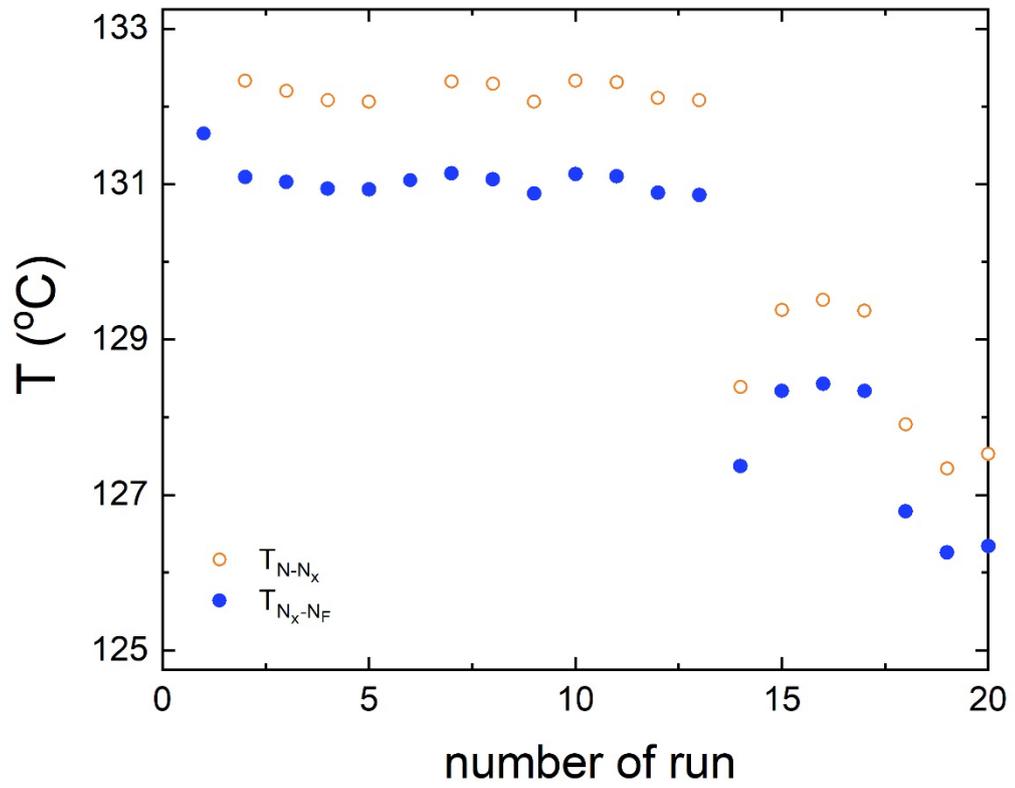

FIG. 6

Calorimetric evidence for the existence of an intermediate phase between the ferroelectric nematic phase and the nematic phase in the liquid crystal RM734

______________________________________________________________________

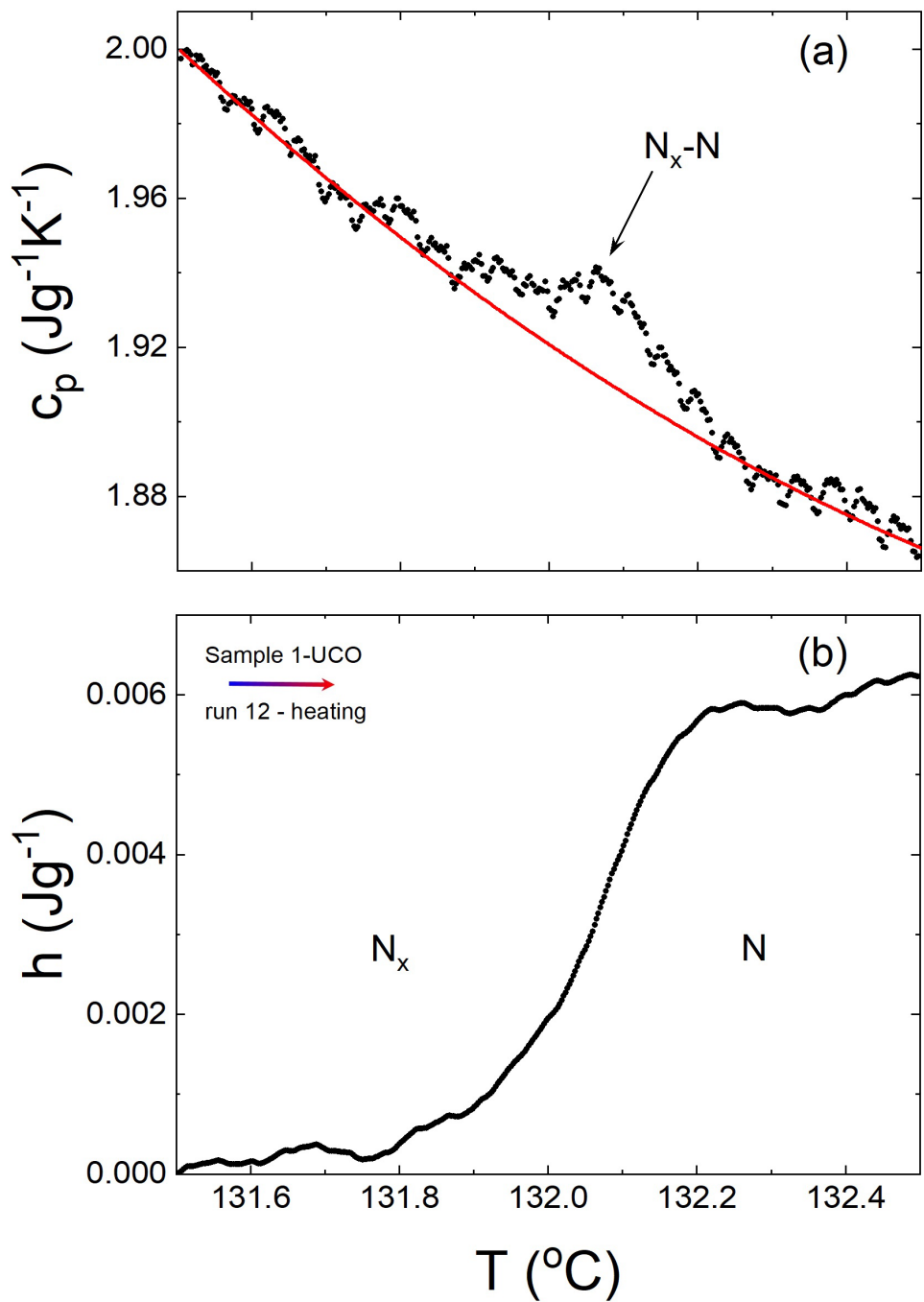

FIG. 7

Calorimetric evidence for the existence of an intermediate phase between the ferroelectric nematic phase and the nematic phase in the liquid crystal RM734

___